\documentclass[preprint,showpacs,preprintnumbers,amsmath,amssymb,showkeys]{revtex4}

\usepackage{graphicx}
\usepackage{dcolumn}
\usepackage{bm}
\usepackage{mathrsfs}
\usepackage{amsmath}
\usepackage{enumerate}

\begin{document}

\preprint{APS/123-QED}

\title{Determining the filling factors of fractional quantum Hall states using knot theory}

\author{C.  Pe\~na}
\email[carlos.andres.pena.castaneda@pwr.wroc.pl]{}
\affiliation{%
Institute of Physics, Wroc{\l}aw University of Technology,
Wyb. Wyspia{\'n}skiego 27, 50-370 Wrocław, Poland
}%


\date{\today}

\begin{abstract}
In this work a method based on a topological invariance of rational tangles commonly used in 
knot theory determines filling factors in the fractional quantum Hall effect. The main sustain for this hypothesis are 
the Schubert's theorems  which treats the isotopic between two knots that are numerators of non-equivalent rational tangles.
This isotopic allows to deduce a new formula for all filling factors. Besides, it opens a new perspective for a future connection 
between  $N-$particles interaction at different fillings and Berry phase evaluated along torus knots. 
\end{abstract}

\pacs{71.10.Pm, 73.43.Cd, 03.65.Fd, 73.43.-f}
\keywords{Quantum Hall effects, braid groups, knot theory}
\maketitle

\section{\label{sec:level1}Introduction\protect\\}

The first sign for fractional Hall resistance was discovered  by Tsui \textit{et al}  in 1982 \cite{Tsui1982, Stormer1999}. 
A high longitudinal conductivity was observed in GaAs-AlGaAs accompanied with a plateu in the Hall resistance within a 
filling factor of $1/3$. The number of filled Landau levels (LLs)  characterises the filling factor as
$\nu=\frac{\rho h c}{eB}$, where the electron density $\rho$, and magnetic field $B$,  determine the  Hall resistance
as $R=\frac{h}{\nu e^2}$. One year later Laughlin 
published a $N-$particles wave function for the lowest Landau level (LLL) associated with fillings $1/q$ with $q-$odd
\cite{LaughlinPRL1983}. The comparison with numerical calculations showed Laughlin wave function was highly accurate,
for a discussion see  ref. \cite{Yoshioka01041985}. 
Since the fraction $\nu$ is a dimensionless parameter one may attempt to define it as 
$\nu=p/q$ with $p$ and $q$ relatively primes. The case $q=1$ is known as the integer quantum Hall effect (IQHE). 
It was discovered by Klitzing in 1980 \cite{Klitzing1980}. Its explanation consists
in the  interaction between an electron and the potential vector. The last is  produced from the strong magnetic field 
applied perpendicular to a two dimensional (2D) sample where  $N-$electrons 
remain confined at a very low temperature 
about 10mK \cite{JainBook2007,Heinonen1998}. In contrast, the fractional quantum Hall effect (FQHE) which occurs 
at $q\ne 1$ is still partially understood. 
It is belived to be originated from the Coulomb interaction between electrons [\cite{JainBook2007}, p.4]. 
A consistent description of the physical mechanism of FQHE ought to  
identify its buildding blokcs, explain incompressibility  \cite{Jain2014} and determine its state of matter at 
specific filling factor. One of the way to solve this enigma is by a construction of 
$N-$particles wave function. Thus it sHall be able to answer not only why the filling factors is a 
fractional number but to capture other fundamental feautures such as spin polarization and topological invariants. 
The last is the main topic of this paper. Furthermore an explanation of all Landau fillings with even denominator like the fraction $5/2$. 
The most popular hypothesis for solving this  puzzle are wonderfull presented in  \cite{Jain2014}. One of them is the composite fermions (CF) approach created by Jain \cite{JainPRL1989}. 
In this model the FQHE is produced by CF which are electrons within an even number of 
quantized fluxes. In CF theory the filling factor is   
\begin{equation}\label{nuCF}
 \nu=\frac{\nu^*}{2n \nu^* \pm 1}
\end{equation}
with $n = 0,1,2,3,...$ 
The trial wave function of CF at the LLL  is  $\psi_\nu=P_{LLL}\prod\limits_{i<j}^N (z_i-z_j)^{2n} \phi_{\nu^*}$, 
where $\phi_{\nu^*}$ is the wave function of noninteracting electrons at integer filling $\nu^*= p$ \cite{JainBook2007}. 
The idea behind this theoretical concept is that FQHE of electrons is created from an IQHE of CF. A successful comparison with experimental data and
numerical calculations of the exact Coulomb energy \cite{WojsPRL2014,Jain2014} has given to CF theory an extensive advantange over other models
\cite{Pfeiffer1993}. However certain fractions like $5/2$, $4/11$, $5/13$  do not fix into 
 formula \eqref{nuCF}. In fact the fraction $5/13$ has been bautized as unconventional \cite{WojsPRL2014}. 
Nevertheless they can be obtained with other expressions (\cite{JainBook2007}, p. 207) 
than certainly are different than  \eqref{nuCF}. The other good candidate to explore  is the hierarchy theory 
of Halperin and Haldane (HH) \cite{Haldane1983,HalperinPRL1984, Yoshioka01041985}. In the HH hierarchy a 
daughter state generates Laughlin quasiparticles   
with fillings calculated from the continued fraction  $\nu = [0, m, \pm q_2, \pm q_3, \pm q_4,...,\pm q_n]$,
where $m=3$ and $q_n=2$. The HH model  is a
hierarchy of fractions started from $1/m$. It reproduces all filling factors with odd denominators. 
The FQHE is explained by  interplay of Laughlin quasiparticles at filling $1/m$ with fractional charge and fractional braid statistics.
This braid statistics  generates daughter states at filling $\nu$ \cite{Jain2014}. The physical assumptions behind CF and HH models arrange filling
factors followed by a construction of wave fucntions of exotic particles with different charge and mass than electrons. 
Even a more amazing conclusion  is that for a given filling factor corresponds
a unique wave function. So, why $\nu$ is a exclusive parameter of the FQHE? The answer perhaps concerns with a relation 
between topology and quantum physics. In 80's Thouless and  Kohmoto \cite{Thouless1985,Kohmoto1985} have discovered that conductivity determines the integer filling
connected with the winding number of a closed curved hooking a torus in a parameter space.  This interpretation works nice for 
a topological description of the IQHE. For the FQHE, conductivity was  produced by including degenaracy of 
the LLL. In 2006 Kohmoto \textit{et al} \cite{WuPRL2006} had calculated the Kubo conductivity from ideas based on gauge invariance. 
The particle statistics was implemented from a braid group formalism on a torus. Although the CF and HH approahes are quite useful as a starting point for description 
of FQHE  they still can not account in a unique way for
wave functions at a filling with even denominator. In addition an explanation
for the recent experimental results \cite{Dehghani2013, Willett2013} 
in connection with the overabundance of filling factors in lower LLs relative 
to thouse with higher LLs. On the other hand it is known that  statistics in 2D systems 
is rather anyons than only fermions or bosons since the fundamental group of the configuration space is isomorphic 
to the braid group $B_n$ \cite{WuPRL1984}. Progress in this area has been done within the cyclotron braid approach by Jacak \textit{et al} in \cite{Jacak2014}. A braid group called the cyclotron braid group with generators 
$b^{(n)}_i=\sigma^n_i$ has been introduced  
in references \cite{Jacak2010, Jacak2011, Jacak2011ATMP, Jacak2012}. This is an alternative framework founded in braid exchange
and cyclotron orbits that describes the exotic statistics of CF allowing a braid interpretation
of Laughlin correlations \cite{Jacak2010}. The geometrical presentation of these braids are 
arcs with an odd number of crossings given by $k$. Filling factors reproduced thouse in \eqref{nuCF} by the expression
\begin{equation}\label{nuJacak}
 \nu=\frac{p}{p(k-1)\pm 1}~.
\end{equation}
In this paper we want to motivate an application of knot theory for a future construction of a quantum and topological description of 
CF's. The main message from braids is a codification of anyon statistics by exchange of coordinates in the $N-$particles wave function \cite{Wilczek1989,Wilczek1990}. Therefore
is is natural to introduce the very well known connection between braids and knots. For instance the Alexander's theorem in \cite{Murasugi1996} 
asserts that every closed braid is isotopic or in other words is topological equivalent to a knot. Therefore for every braid interpretation of CF there is an asociated knot 
description. In this work the richness of knot theory will be suitable for deduction of a formula from which is possible to obtain 
all filling fractions of CF including thouse with even denominator. The formula is deduced for a straightforward application of 
Schubert's theorems for isotopic of  knots numerators extracted from rational tangles \cite{RiccaI2001,Kauffman2003, Kauffman2004}. 
This theorem was originally formulated 
for closed braids named 4-plats \cite{Schubert1954}, (\cite{Birman1974}, pp. 212) or two bridge knots, see for instance ref. \cite{Kawauchi1996, Burde1985}. The Schubert's theorems
for rational knots were analized by Kauffman \textit{et al} in ref. \cite{Goldman1997,Kauffman2003,Kauffman2004}. By following the 
structure of these theorems one may deduce that every filling factor is a topological invariant of rational 
tangles. This amazing fact is known as the  Conway's theorem mentioned in \cite{Kauffman2003} and proven in ref.
\cite{Burde1985, Goldman1997, Kauffman2004}. Rational tangles were introduced by Conway in 1967  within a geometrical presentation 
\cite{Conway1967}. Every rational tangle composed with two strands has a 3-braids representation \cite{Kauffman2003,Kauffman2004}. Additionally 
every rational knot has a 4-plat presentation and is formed by closing a tangle \cite{Birman1974, Burde1985, Kawauchi1996, Goldman1997}. Moreover the theory of classical vortices has been formulated in terms of knots \cite{RiccaI2001, RiccaII2001}. Some progress with 
quantum vortices such as thouse in superconductivity can be inflicted from a Chern-Simons theory \cite{Enger1998,AnaPRB1991,Zhang1992,Kossow2009}. Amazingly 
path integrals of a Chern-Simons theory has explicitly shown the tipical skein relations of Jones polynomials which 
are invariants of knots \cite{Witten1989,Witten2013,Natan1995}. The partition function defined from a Boltzmann factor has been used toghether with the 
Yang Baster equation to recovery Jones polinomials \cite{WuRMP1992,Murasugi1996}.  This connection can be profitable for characterization of 
topological properties of systems such as topological insulators \cite{HasanRMP2010}, distribution functions of 2D particle systems  \cite{Lado2003}, 
FQHE in graphene \cite{Jacak2012, HaldanePRL1988} and Wigner crystals  \cite{Kivelson1986}. In this work we will see that 
the Schubert's theorems accomodate for a determination of filling factors for CF. The application is justified by the Conway's theorem 
where filling factors can be taken as fractions that are topological invariants of rational tangles. 
Furthermore, the the Schubert's theorems for rational knots, obtained as knot numerators of rational tangles, endure to establish a 
classification of filling factors of CF via isotopic of rational knots. Additionally an explanation of why filling factors 
with high numerator are rare can arise in knot theory
from a relation between conductivity and Berry phase evaluated along torus knots \cite{RiccaII2001}. Later the electric current is calculated by using 
a method based on a Laughlin's idea explained in ref. \cite{LaughlinPRB1981}. The result suggests 
an explanation of rare fillings due to innestabilities of the transformed potential energy caused by the Berry phase. 

In Section \ref{section:tangles} we begin with the definition  of rational tangles and isotopy of knots. 
Besides a tangle method for filling fractions is described.
In Section \ref{section:formula} the Schubert's theorem for unoriented tangles determines a
formula for filling factors of FQHE. The formula is generalized for the case of oriented tangles.
In Section \ref{section:conductivity}, a physical interpretation of a tangle is given in terms of conductivity. This 
interpretation is based
on the fractional decomposition of 
the product of a rational winding number for torus knots 
and the Chern number. In Section \ref{section:energy}, a special case of planar isotopy 
is applied to the Hamiltonian. This isotopy transformation which is defined by a deformation of knots,
yields a relation between the Coulomb energy and filling factors.

\section{\label{sec:level2}Rational Tangles}\label{section:tangles}

Tangles were introduced originally by Conway in 1967  with the purpose of
classifying alternating knots. A tangle is defined as a portion of a knot diagram from which there emerges just 4 arcs pointing
in its compass direction \cite{Conway1967}. Examples of how tangles look like can be seen in ref. \cite{Conway1967, Kauffman2004}.
A special case of tangles are rational tangles which do not contain separated loops but only alternatings twist of 
two strands. An extension of these  definition to arbitrary number of strands is likewise possible. A more formal definition
of tangles in terms of homeomorphism is given by Kauffman and Lambropoulou in ref. \cite{Kauffman2004}.  It is well known  in knot theory 
\cite{Kauffman2003,Murasugi1996, Burde1985, Kawauchi1996,Watkins1990}
that a topological invariant of a rational tangle $T$ is a fraction   
denoted as F(T). Every fraction defined with relatively primes numbers can be decomposed in a continued fraction \cite{Olds1963, Burde1985}
defined by the expression 
\begin{eqnarray}\label{fraction} 
  F[T] &=& [a_1, a_2, a_3,...,a_m]
\nonumber\\
&=&a_1+\cfrac{1}{a_2
          + \cfrac{1}{a_3
           +\cfrac{1}{..+\cfrac{1}{a_k}} } }
\end{eqnarray}
if $k$ is odd and either all $a_i>0$ or $a_i<0$  we say the fraction is written in a canonical form \cite{Kauffman2003,Kauffman2004}. 
For $k$-even it is easy to transform the fraction in a canonical form since $[a_1, a_2, a_3,...,a_k]=[a_1, a_2, a_3,...,a_k-1,1]$.
In fact every continued fraction can be transformed to a unique canonical form. The associated tangle in statndard form
\cite{Conway1967,Kauffman2003}  is defined as 
\begin{eqnarray}\label{tangle} 
  T &=& [[a_1], [a_2], [a_3],...,[a_m]]
\nonumber\\
&=& \biggl(\Bigl(([a_m]_o + [a_{m-1}])_o + ... + [a_3]\Bigl)_o+[a_2]\biggl)_o+ [a_1]
\nonumber\\
\end{eqnarray}
within a topological invariant  given by  \eqref{fraction} with properties
 \begin{subequations}
\begin{eqnarray}
&a.& F(T+ [\pm 1]) = F(T) + 1~,\label{f1}
\\
&b.& F\Bigl(\frac{1}{T}\Bigl) = \frac{1}{F(T)}~.\label{f2}  
\\
&c.&  F(-T) = -F(T)~.\label{f3}         
\end{eqnarray}
\end{subequations}

\begin{figure}[!htbp]
\includegraphics[height=0.6in]{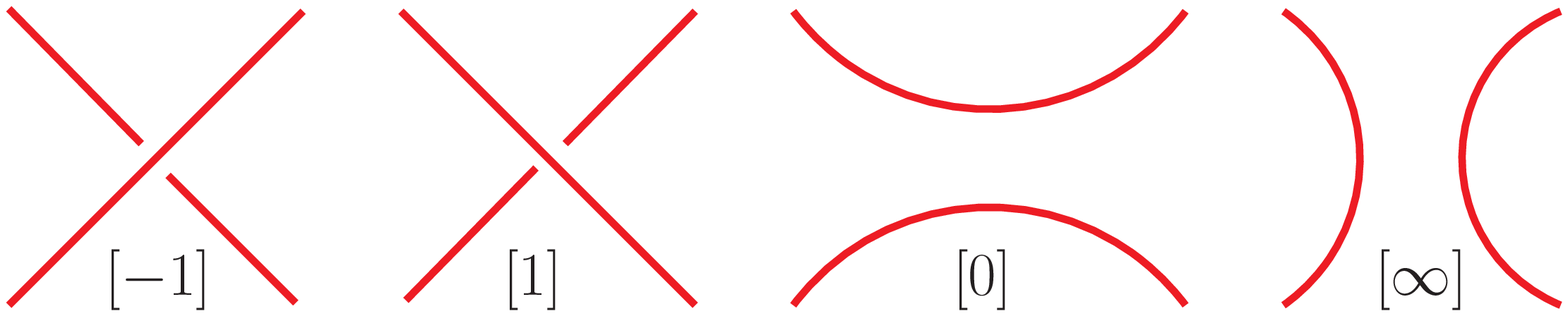}
\caption{\label{fig1} All tangles are composed of these elementary tangles \cite{Conway1967}.}
\end{figure}

A soft way to understand this definition is using a diagramatical reresentation of tangles.
One may start knowing that all tangle diagrams are composed of the  fundamental structures
as in Fig.~\ref{fig1}.

\begin{figure}[!htbp]
\includegraphics[height=0.9in]{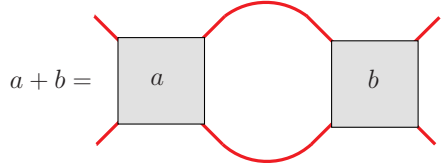}
\caption{\label{fig2} Sum of two tangles. Just join the adjacent arcs.}
\end{figure}

\begin{figure}[!htbp]
\includegraphics[height=0.9in]{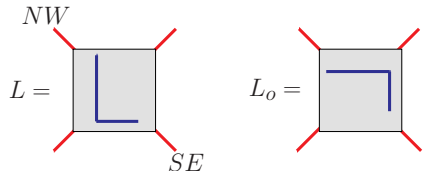}
\caption{\label{fig3} $L_0$ is the result of reflecting $L$ through its principal diagonal in the compass direction NW-SE \cite{Conway1967}. This
introduce the Conway's product for tangles as $ab=a_0+b$.}
\end{figure}

\begin{figure}[!htbp]
\includegraphics[height=0.8in]{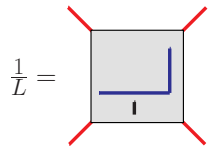}
\caption{\label{fig4} Inversion of a tangle \cite{Kauffman2003, Kauffman2004}. Rotating the tangle counterclock-wise
by $90^\circ$ form a tangle $L^r$. The mirror image of a tangle is formed by switching all the crossings $-L=-(L)$. The
inversion of a tangle is this rotation followed by its mirrow image.}
\end{figure}

The  sum of two tangles $a+b$ is illustrated in Fig.~\ref{fig2}. A reflexion  $L_o$ in  Fig.~\ref{fig3}.
and the inversion are defined by  Fig.~\ref{fig4}.

Now a very important issue is the isoptopy of rational tangles \cite{Kauffman2003, Murasugi1996}. It  was established in 1975 
as \textbf{Theorem 1 (Conway):} \emph{Two rational tangles are isotopic if and only
if they have the same fraction.} What it basically means is that a fraction is a topological invariant of rational tangles
which are topological equivalent or in other words isotopic \cite{Conway1967, Kauffman2003, Burde1985}.

\begin{figure}[!htbp]
\includegraphics[height=1.7in]{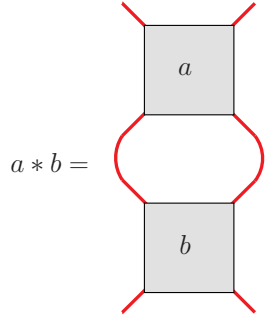}
\caption{\label{fig5} Kauffman product is a binary product. Moreover 
with this product the rational tangle in \eqref{tangle} can be as well defined \cite{Kauffman2003}.}
\end{figure}

The Conway's product of two tangles is a binary product defined as $ab=a_o+b$ \cite{Conway1967}. The Kauffman product 
$a*b$ \cite{Kauffman2004} is as in Fig.~\ref{fig5}

\subsection{\label{sec:level3}Knot Numerator of a Rational Tangle}

\begin{figure}[!htbp]
\includegraphics[height=1.3in]{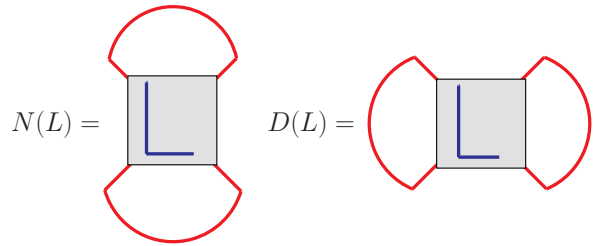}
\caption{\label{fig6} Knot numerator and denominator of a tangle $[L]$. We denote the numerator of a tangle $L$ as  $N(L)$ 
and denominator as $D(L)$.}
\end{figure}

The connection between tangles and knots are acomplished by the closure of tangles as in  Fig.~\ref{fig6}.

Here the knot numerator is a rational knot. Every rational knot is an alternating  knot \cite{Kauffman2003}. 
The Schubert's theorems deals with isotopy of knot numerators. Before going into this theorem we briefly review the idea of isotopy.

\subsection{Concept of Isotopy}

Isotopy between two knots means intuitively that a given knot can be deformed continuously into other by a surjective homotopy.
Then one say the two knots are topological equivalent \cite{Kawauchi1996, Murasugi1996, Burde1985}.
Therefore if two knots are isoptopic they define the same knot.
Isotopy is formally express as:  
\emph{Let be  $X$ and $Y$ two topological spaces and $f:X \to Y$ and $g:X \to Y$ homeomorphims. 
The functions f and g are isotopic and we denote $f\sim~g$ iff there exists a continuous surjective homotopy 
defined by the function $H(x,t)$ such that
\begin{eqnarray} 
H: X \times [0,1] \to Y 
\end{eqnarray}
with $H(x,0)=f(x)$ and $H(x,1)=g(x)$.}

\begin{figure}[!htbp]
\includegraphics[height=3.2in]{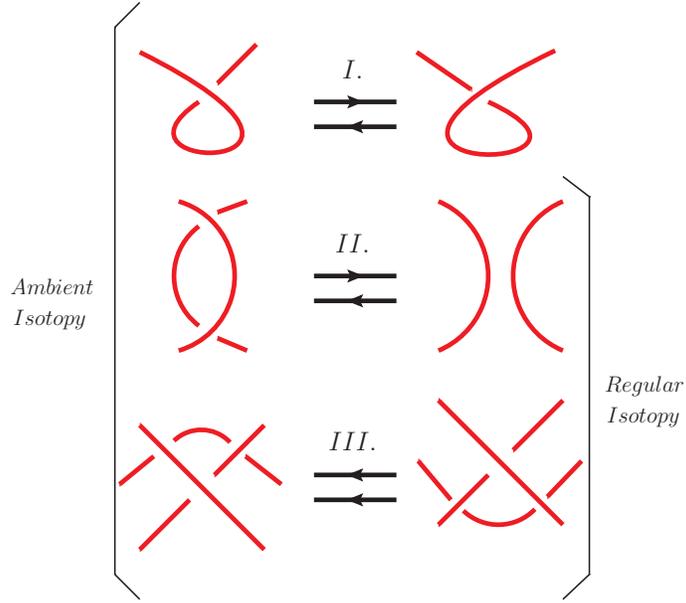}
\caption{\label{fig9} Reidemeister moves of type \emph{I}, \emph{II} and \emph{III}. These moves can be performed in 2D 
on a regular diagram of a knot.}
\end{figure}

There is an more efficient definition of isotopy for knots. It was originally proven the equivalence to the former 
in terms of Reidemeister moves  \cite{Reidemeister1927}. More recent proofs in (ref. \cite{Burde1985} p.1-9, 193), (ref. \cite{Kawauchi1996} p.4, appendix A) 
(ref. \cite{Murasugi1996} p.50, 13).  The Reidemeister moves are transformations  performed on parts of a knot.
So take a knot and look for any of the configurations depicted in Fig.~\ref{fig9}, then for that configuration it is allowed to 
do a Reidemeister move. There are three Reidemeister move  described as following  \cite{Kauffman1990}
\begin{enumerate}[I.]
\item Twist and untwist a curl in either direction,
\item Lie down one loop completely over another,
\item A string can be moved over or under a crossing.
\end{enumerate}
If two knots are related through a finite number of  Reidemeister moves of type I, II, III
they are said ambient isotopic \cite{Murasugi1996, Burde1985, Kauffman1990}. Similarly 
they are regular isotopic if one can be transformed into another by the moves of type II and III. 
Other definitions includes plane isotopy  \cite{Web}
as deformations  \cite{Murasugi1996} such as  rotation around a fix point, drag, 
translation and dilatation/elongation.

\subsection{\label{sec:level3}The Tangle Method}\label{section:method}

\begin{figure}[!htbp]
\includegraphics[height=3.7in]{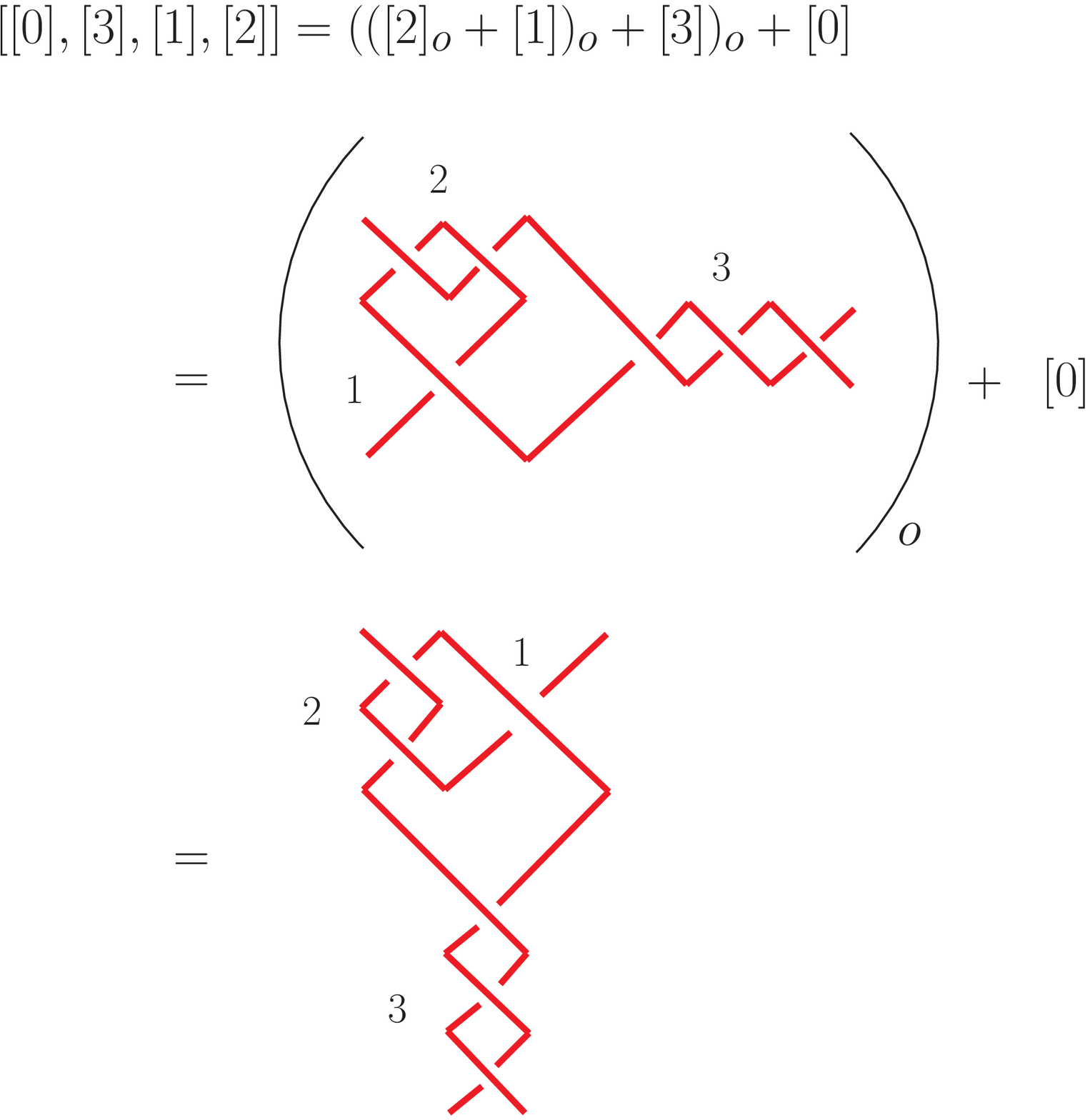}
\caption{\label{fig7} Presentation of a tangle with  invariant fraction $\frac{3}{11}$. The knot numerator of this tangle is isotopic to 
the knot $3_1$.}
\end{figure}

One may use tangles to classify the filling factors of FQHE. 
Let select a fraction in of FQHE, for intance $\frac{3}{11}$. It can be written as $[0,3,1,2]$
which is equals to a continued fraction in canonical form given by  $[0,3,1,2-1,1]=[0,3,1,1,1]$.  
The tangle whose topological
invariant is the fraction $\frac{3}{11}$ can be written in the standard form as $[[0],[3],[1],[2]]$. In order to find
its  knot numerator a presentation of the tangle is required. Following the expression \eqref{tangle} 
and the basic operations given in Fig.~\ref{fig1},  Fig.~\ref{fig2},  Fig.~\ref{fig3} 
one may find easily the result in Fig.~\ref{fig7}.
The numerator closure of this tangle is isotopic to the knot $3_1$. Tables of knots up to twelve crossings can be consulted in
the Rolsen table and the Hoste-Thistlethwaite table \cite{Atlas}. A full table of 
knots with twelve crossings can be seen in \cite{TableKnots}.
For more information, we address the reader with a didactical explanation 
in references \cite{Conway1967,Kauffman2003, Kauffman2004,Goldman1997}.

\subsection{\label{sec:level3}Recovery the Cyclotron Group }

\begin{figure}[!htbp]
\includegraphics[height=1.3in]{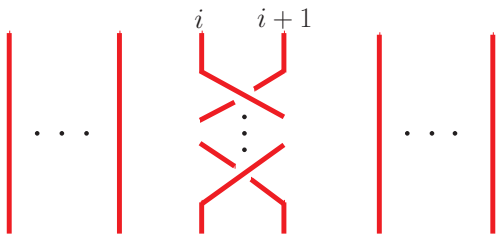}
\caption{\label{fig8} Presentation of the braid $\sigma^k_i$ with $i=1,2,3, ... N-1$.}
\end{figure}

One may recovery  the cyclotron braid group generators introduced in \cite{Jacak2012} by 
using a $(N-1)-$Tangle. Therefore it is easy to see in Fig.~\ref{fig8} that if $k-$odd the tangle $[[0],[k]] = [k]_o + [0]$  
is  equals to $\sigma^k_i$.

It gives exactly the generator $b^{(k)}_i$ introduced in reference (\cite{Jacak2012}, p.27-29).

\subsection{Schubert's Theorem for Unoriented knots}

In this section is extressed that filling factors
of FQHE are topological ivariants of rational tangles and can be organized by isotopy of rational knots. 
Here we present a theoretical  support.
The  Schubert's theorems for unoriented/oriented knots were  introduced
using 4-plats presentations of knots \cite{Schubert1954, Kawauchi1996, Burde1985, Birman1974}. 
Developments with tangles are found in references \cite{Kauffman2003, Kauffman2004, RiccaII2001}. The first 
theorem is\\

\textbf{Theorem 2 (Schubert):}\emph{
Suppose that rational tangles with fractions
$\nu=\frac{p}{q}$ and $\nu'=\frac{p'}{q'}$ are given ($p$ and $q$ are relatively prime. Similarly for $p'$ and $q'$.) If
$N[\frac{p}{q}]$ and $N[\frac{p'}{q'}]$ denote the corresponding rational knots obtained by taking
numerator closures of these tangles, then $N[\frac{p}{q}]$ and $N[\frac{p'}{q'}]$ are isotopic if and
only if}

\begin{enumerate}
\item\label{1}  \emph{$p = p'$ and }
 \item\label{2}  \emph{either $q =\pm q'+ m p$ or $qq'= \pm 1 + m p$}. 
\end{enumerate}
\emph{with m-integer.} Since a knot is isotopic to itself one should take $m=1,2,3,...$

\section{A Formula for filling Factors in FQHE}\label{section:formula}

The idea that filling factors are topological invariants was used in 1985 by Thoules \textit{et al} \cite{Thouless1985}
via a calculation of conductivity. In this paper the Conway's theorem is applied since it admits an interpretation of
filling factors in FQHE as a topological invariants of rational tangles.  A relation
between tangles and conductivity will be appointed in Section \ref{section:conductivity}. 

The condition $p=p'$ and $q =\pm q'+ m p$ of Thorem 2 is sumarized in the Kauffman's equation \cite{Kauffman2004}
\begin{eqnarray}\label{kauffmanEq} 
(\pm T)*\frac{1}{[m]} = \frac{1}{[m]\pm\frac{1}{T}}~.
\end{eqnarray}
The equation  \eqref{tangle} can be obtained inductively from \eqref{kauffmanEq}. 
In any case if we define $F((\pm T)*\frac{1}{[m]}) = \frac{p}{q}$ and $F[T] = \frac{p}{q'}$
is easy to apply the propierties of   \eqref{f1}-\eqref{f3} hence 
\begin{eqnarray}\label{FractionEq} 
F\left((\pm T)*\frac{1}{[m]}\right)&=&\frac{1}{m \pm \frac{1}{F(T)}}~,
\end{eqnarray}
and then 
\begin{eqnarray}\label{formula1} 
\frac{p}{q} &=& \frac{p}{ m p \pm q'}~.
\end{eqnarray}
Certainly this result can be inferred directly from Schubert's theorem. In terms of filling factors  
the formula \eqref{formula1} transforms in
\begin{eqnarray}\label{formula11} 
\nu = \frac{\nu'}{ m \nu' \pm 1}~.
\end{eqnarray}
This formula  is presented
including a formal concept of topological invariance and can be used to reproduce conventional filling
factors in the FQHE. Notice that this theorem does not imposse
restrictions on the integer $m$. Nevertheless it is effortless to see the Jain's formula
\eqref{nuCF} is a special case of \eqref{formula11} when  $m=2n$. The cyclotron formula \eqref{nuJacak}
is the case when  $m=k-1$. Additionally many filling factors  of the HH theory are obtained
by  \eqref{formula11}. At this point it is worth to mention that the fraction decomposition
of the HH hierarchy is non-canonical since not all $a_i$ in \eqref{fraction} would have the same sign. One may remember that
a canonical form of a fraction leads to rational tangles that are alternating. For instance, a canonical
form of the fraction $\frac{4}{9}$  is $[0,2,4]$ which is the invariant of a rational tangle
whose numerator clousure is the alternating link $L4a1$. 
In contrast,  this fraction in the HH model is decomposed as 
$\frac{4}{9} = [0,3,2,-2,-2]$ yielding a no-rational tangle. Since in the 
HH theory the components of a continued fraction are fixed by the hierarchy
there is not freedom to organize them via the isotopy of alternating knots.

\subsection{Conventional versus Unconventional FQHE}\label{ConvenUnconventional}

The definition of unconventional fillings was introduced by   W\'ojs \textit{et al} in \cite{WojsPRL2014}. 
Fractions which are out of expression \eqref{nuCF} were defined as unconventional.
Is there any topological reason for that?  For instance, in Table~\ref{tab:table7} the 
invariants $\frac{7}{2}$, $\frac{7}{3}$, $\frac{7}{5}$ can not be obtained 
just by replacing in \eqref{formula11} the filling factor of IQHE given by $\nu'=7$. So
one may call them unconventional. A topological description can be constracted 
observing that the fraction $\frac{7}{2}$ is an invariant of the tangle
$[[3],[2]]$ which has a knot numerator isotopic to $5_2$ which does not belong to the Table~\ref{tab:table0}.  In contrast  
$\frac{7}{13}$, $\frac{7}{15}$, $\frac{7}{27}$ are calculated using \eqref{formula11} having knot numerators as
$7_1$. This is  the same as the knot numerator of tangle with invariant 
$7$. This last fraction is associated to the seventh Landau level in the IQHE regime. 
Similarly in Table~\ref{tab:table5} the tangle 
whose invariant is the fraction $\frac{5}{11}$ has as a  knot numerator $5_1$, the same as the knot numerator of the tangle with invariant 
$5$ which associates with the fifth Landau level in the IQHE. This is not the case of the 
tangle with invariant $\frac{5}{2}$ with knot numerator rather isotopic to $4_1$. As we may observe  $4_1$ is not part of the invariants 
connected to the IQHE. Therefore  in this context $\frac{5}{2}$ is unconventional as well
and can not be obtained with formula  \eqref{formula11}. In topological terms, one may say that a
filling factor for the FQHE  is conventional if it is a topological invariant of 
a rational tangle with  knot numerator isotopy to one of the IQHE. Otherwise is 
unconventional.

Consequently it would be convenient to construct a unique equation for conventional as well as unconventional 
fillings. Firstly let us notice  that \eqref{kauffmanEq} and
\eqref{formula11} do not bear enterely the second option  impossed by Theorem 2. It is allow to 
have as well $qq'= \pm 1 + m p$. Therefore applying the full theorem 2  one obtain two options

\begin{equation}\label{options}
\nu = \begin{cases} 
       \frac{p}{ m p \pm q'}      & \text{for ~ $q' =\pm~q'+ m p$}~,\\
       \frac{p}{\frac{m}{q'}p \pm \frac{1}{q'}} & \text{for ~ $qq'= \pm~1 + m p$}~.
        \end{cases}
\end{equation}
One may define  $q=\alpha p~\pm~Q$ 
where  either  $\alpha=m$ and $Q=q'$ or  $\alpha=\frac{m}{q'}$ and $Q=\frac{1}{q'}$. 
Then both options in \eqref{options} can be written in a unique equation 
given by
\begin{eqnarray}\label{formula2}
\nu &=& \frac{p}{\alpha p~\pm~Q}~.
\end{eqnarray}
Where $p$ is the filling of the IQHE and $q'$ 
might be interpreted as a parameter for the FQHE constrained to the former definitions. There are no obvious mathematical
arguments which restrict the values of $p$ and $q'$ but there could be physical reasons related with the dynamics of electrons.
An attempt for a physical interpretation of filling factors in terms of rational tangles is given in the next section.
Thus the Kauffman's equation \eqref{kauffmanEq} would be 
\begin{eqnarray}\label{kauffmanEq2} 
(\pm T)*\frac{1}{[\alpha]} = \frac{1}{[\alpha]\pm\frac{1}{T}}~.
\end{eqnarray}
where $[\alpha]$ is a rational tangle constrained to the rule \eqref{tangle}. Additionally
$\nu = F((\pm T)*\frac{1}{[\alpha]})$ and $F(T) = \frac{p}{Q}$. 
Now  filling factors for the FQHE are  organized via isotopy
of knot numerators. The results are in Tables~\ref{tab:table0}-\ref{tab:table10}. Here below it is described 
briefly the procedure. For illustration take  tangles connected with the IQHE region. Let it be $T=[1]$, with fraction
$F(T)=1$ corresponding to the first LL. Therefore $p=1$, $Q=1$ and $\alpha=m$. This generates all fractions of the form 
\begin{equation}
\frac{1}{q}=\Bigl\{1,\frac{1}{2}, \frac{1}{3}, \frac{1}{4}, ...\frac{1}{m \pm 1} ,...\Bigl\}~. 
\end{equation}
which are invariants of the tangles $[[0],[q]]$ with knot numerators isotopic to $0_1$. 
Successively $F(T)=2$ leads to
\begin{equation}
\frac{2}{q}=\Bigl\{2,\frac{2}{3}, \frac{2}{5}, \frac{2}{7}, ...\frac{2}{2m \pm 1} ,...\Bigl\}~. 
\end{equation}
with knot numerator isotopy to the link $L2a1$. $F(T)=3$ yields
\begin{equation}
\frac{3}{q}=\Bigl\{3,\frac{3}{2}, \frac{3}{4}, \frac{3}{5}, ...\frac{3}{3m \pm 1} ,...\Bigl\}~. 
\end{equation}
with knot numerator isotopy to the link $3_1$. $F(T)=4$ generates
\begin{equation}
\frac{4}{q}=\Bigl\{4,\frac{4}{3}, \frac{4}{5}, \frac{4}{7}, ...\frac{4}{4m \pm 1} ,...\Bigl\}~. 
\end{equation}
with knot numerator isotopy to the link $L4a1$.  $F(T)=5$  leads to
\begin{equation}
\frac{5}{q}=\Bigl\{5,\frac{5}{4}, \frac{5}{6}, \frac{5}{9}, ...\frac{5}{5m \pm 1} ,...\Bigl\}~. 
\end{equation}
Amazingly  the unconventional fillings as $\frac{5}{2}$, $\frac{5}{3}$, $\frac{5}{7}$, $\frac{5}{13}$
 in Table~\ref{tab:table5}  are associated to the knot $4_1$ which is 
not isotopic to any of the knot numerators for the IQHE. Hence one should have a new starting point 
with the filling factor of $F(T) = \frac{5}{2}$ and then  $\alpha = \frac{m}{2}$, $Q=\frac{1}{2}$  yields
\begin{equation}
\Bigl\{\frac{5}{2}, \frac{5}{3}, \frac{5}{7}, ...\frac{10}{5m \pm 1} ,...\Bigl\}~,
\end{equation}
here  $2q = 5m \pm 1$  is an even number and so for unconventional fillings originated from  $F(T) = \frac{5}{2}$ not all values of $m$ are allowed.  
Thus e.g. fillings like $\frac{10}{9}, \frac{10}{11}$ do not belong to the same clasification since $5(2)- 1=9$ and 
$5(2) + 1=11$ are odd denominators. 
 
Similarly there are unconventional fillings in Tables~\ref{tab:table7} and \ref{tab:table9}.

\subsection{Shubert's Theorem for Oriented Knots}

Theorem 2 deals with isotopy of unoriented rational knots constructed from unoriented tangles. 
This theorem changes for the case of oriented knots. Furthermore if two tangles have different orientation 
their corresponding knot numerators might be distint
even if the invariant fractions are the same. Therefore knots numerators must be built  
from compatible-oriented tangles. Two tangles are compatible-oriented if their end arcs have the same orientation.  
Notice that a tangle $[T]$ is orientation-compatible to $T*[2n]$ consequently $N(T)$ is isotopic to 
$N(T*[2n])~$ (\cite{Kauffman2003}, p. 45). The
Theorem 3  elucidates a formula for invariants fillings of orientable tangles.\\ 

\textbf{Theorem 3 (Schubert):}\emph{ Suppose that orientation-compatible rational tangles with fractions
$\frac{p}{q}$ and $\frac{p'}{q'}$ are given ($p$ and $q$ are relatively prime. Similarly for $p'$ and $q'$.) If
$N[\frac{p}{q}]$ and $N[\frac{p'}{q'}]$ denote the corresponding rational knots obtained by taking
numerator closures of these tangles, then $N[\frac{p}{q}]$ and $N[\frac{p'}{q'}]$ are isotopic if and
only if}

\begin{enumerate}
\item  \emph{$p = p'$ and }
 \item  \emph{either $q =\pm q'+ 2n p$ or $qq' = \pm 1 + 2n p$}. 
\end{enumerate}
\emph{with n-integer.}
Thus a general formula for FQHE deduced on the base of Theorem 3 is
\begin{equation}\label{formula3}
\nu = \frac{p}{2\alpha p \pm Q}~,
\end{equation}
with  $q=2\alpha p~\pm~Q$ 
where  either  $\alpha=m$ and $Q=q'$ or  $\alpha=\frac{m}{q'}$ and $Q=\frac{1}{q'}$. 
In both cases the Kauffman's equation for orientable tangles is 
\begin{eqnarray}\label{kauffmanEq3} 
(\pm T)*\frac{1}{[2\alpha]} = \frac{1}{[2\alpha]\pm\frac{1}{T}}~.
\end{eqnarray}
where once more $[\alpha]$ is a rational tangle constrained to the rule \eqref{tangle}.
Let us take once again the tangle $T=[1]$ with fraction
$F(T)=1$ corresponding to the  LLL. This generates  fractions of the form 
\begin{equation}\label{O1}
\frac{1}{q}=\Bigl\{1, \frac{1}{3}, \frac{1}{5}, ...\frac{1}{2m \pm 1} ,...\Bigl\}~. 
\end{equation}
just fillings with odd denominator. Nevertheless for $F(T)=\frac{1}{2}$, $p=1$, $\alpha=m$
and $Q=2$ is obtained the set
\begin{equation}\label{O2}
\Bigl\{\frac{1}{2}, \frac{1}{4}, ...\frac{1}{2m \pm 2} ,...\Bigl\}~. 
\end{equation}
which are fractions with even denominator. However notice that $F(T)=\frac{1}{2}$ has not been generated from the LLL. 
This is a consequence of the orientation of knots numerators. Since the only difference between 
the knots numerators associated to \eqref{O1} and \eqref{O2} is the orientation then a knot numerator corresponding to a rational tangle
with invariant filling factor of odd denominator has opposite orientation with respect to that of a filling of even denominator. 
The orientation of knots can be set experimentally by changing the orientation of the applied voltage applied to the 2D Hall system. 
This hypothesis might be related with the experimental anysotropy announced in ref. \cite{Lilly1999}.  

\section{\label{sec:level3}Where to find Knots in FQHE?}\label{section:model}
In this section we describe two canditates for  knots in the FQHE. 
\emph{First}:  Impossing boundary conditions on a single particle wave function as it was done by Thouless \emph{et al} in ref. 
\cite{Thouless1985} yields a torus which determine a parameter space. For instance a set of lines on the 2D system 
would be seen as a knot on the torus surface. Torus knots have a rational winding number which is a basic ingredient 
of conductivity for the FQHE. See e.g section~\ref{section:conductivity}.
\emph{Second}: It is well known that a voltage must be 
applied to the 2D system in order to observe a current which experience a Hall resistance. One may naively 
imagine a 2-tangle created in time on a 2D sample by the transport of charge carries.
The conductivity for such arrangment would be given by $\frac{e^2}{h} F(T)$. 
Equivalent definitions of conductivity in terms of tangles and knots were given in 
\cite{RiccaI2001,Goldman1993}.

\subsection{Conductivity}\label{section:conductivity}

In this section we will examine the first candidate for a knot in the FQHE. 
The conductivity of a $N-$particles wave function for the FQHE is obtained
by using the continued fraction approach  for rational tangles in terms 
of the product of Chern and winding numbers. The last is associated to a torus knot living on the parameter space. 

We start considering  the average of the Hall conductivity for the N-particles ground state $\psi_k$ on
a 2D square sample of length $L$ with FQHE. A unitary transformation given by $\phi_k=\exp\left[-i\frac{\theta}{L}(x_1+...+x_N)\right] \exp\left[-i\frac{\varphi}{L}(y_1+...+y_N)\right]\psi_k$
was employed by Thouless \textit{et al} within the Kubo's formula \cite{Thouless1985}.
The result including the Berry phase $\gamma_k(\Gamma)$  is
\begin{eqnarray}\label{Conductivity}
 \sigma &=& i\frac{e^2}{h d}\sum\limits^d_{k=1} \int\limits^{2\pi}_0\int\limits^{2\pi}_0
\frac{d\theta\,d\varphi}{2\pi}\left[\left\langle \frac{\partial\phi_k}{\partial\theta} \Bigl| \frac{\partial\phi_k}{\partial\varphi}\right\rangle- 
\left\langle \frac{\partial\phi_k}{\partial\varphi} \Bigl| \frac{\partial\phi_k}{\partial\theta}\right\rangle
\right] 
\nonumber\\
&=& \frac{e^2}{h d}\sum\limits^d_{k=1}\int\limits^{2\pi}_0\int\limits^{2\pi}_0
\frac{d\theta\,d\varphi}{2\pi} \left[\partial_{\theta} A_\varphi-\partial_{\varphi} A_\theta\right]
\nonumber\\
&=&\frac{e^2}{h d} \sum\limits^d_{k=1}\frac{\gamma_k(\Gamma)}{2\pi}~.
\end{eqnarray}
The parameters $\theta,\varphi$ are  phases of a gauge transformation induced by 
magnetic translation operators acting on a single particle wave function, see ref. \cite{Thouless1985}. 
They define a 2D parameter space which can be mapped into a 2D torus due to the boundary conditions
imposed on single particle wave functions. Here in this paper the closed curve on which the Berry phase 
is evaluated is taken as a torus knot living on the 2D torus. In a different approach but still using 
\eqref{Conductivity} Kohmoto \textit{et al}
employed the Kubo's formula in the Brillouin zone \cite{Kohmoto1985} within a periodic potential. 
More reciently the expression \eqref{Conductivity} was used  with braid  and gauge groups on a torus \cite{WuPRL2006}.\\ 

A key point in the FQHE is the ground-state degeneracy  counted by  $d$.
The index $k$ label degeneracy of the ground state.
The relation between conductivity and the Berry phase is well known \cite{Huber2011,Huber2013}.
For a closed curve $\Gamma$, enclosing the parameter space just once,
the Berry phase vanishes modulo $2\pi$ \cite{Sakurai} and the integer Chern number, $C_k$,
determines the conductivity since the Berry phase is
\begin{eqnarray}\label{BerryPhase}
 \gamma_k(\Gamma) &=& \oint\limits_\Gamma A_k\left(R\right)\cdot dR
\nonumber\\
&=&2\pi C_k~,
\end{eqnarray}
and the Berry connection 
\begin{equation}\label{BerryConnection}
 A_k\left(R\right)=i~\langle \phi_k |\nabla_R| \phi_k\rangle~.
\end{equation}
Here $\Gamma$ is a torus knot  drawn 
by the two components dimensionless vector $R=(\theta,\varphi)$.  
Generally  $R$ dependent on time and 
enters in the Hamiltonian (see e.g \cite{Sakurai} pp. 349).
Therefore here $\theta,\varphi$  vary with time and may be introduced in the Hamiltonian as a 
consequence of a gauge transformation \cite{Thouless1985}.  This is a motivation
to define a tangle using the conductivity from a Berry phase as the one in ref. \cite{Huber2013,Thouless1985}.
If the knot given by $\Gamma$  goes around the torus more than one time,
the expression \eqref{Conductivity} is completed including the winding number \cite{Michel2007}  
multiplied by the standard Berry phase \cite{Sakurai}. Then for the FQHE the equation \eqref{Conductivity} provides
\begin{equation}\label{ConductivityTangle}
 \frac{1}{d} \sum\limits^d_{k=1} W_kC_k = F(T)~.
\end{equation}
 We remind by the Conway's theorem that the filling factor  $F(T)=\frac{p}{Q}$ defined in section \ref{ConvenUnconventional},
is an invariant of a rationl tangle. For the IQHE  when $F(T)=p$, all $W_kC_k$  must be equal and integers then degeneracy 
is unimportant since the sum disappear. Then the torus knot 
is  given  by $0_1$ which has a  winding number equals one. 
For the FQHE  the degeneracy of the ground state is relevant.
One may define the total sum on degenerate states as $F(T)=WC$ which is the product of the rational winding number $W$ 
for torus knots and the integer Chern number $C$. In this sense the formula \eqref{ConductivityTangle} which determines 
the conductivity in \eqref{Conductivity} would remains the same 
either for IQHE or FQHE. As we have mentioned the components of a rational tangle correspond to thouse of the continued 
fraction decomposition \eqref{fraction}. Therefore the degree of degeneracy can be
given by the number of components of the rational tangle. For instance, let us take the filling factor
with four components as $\frac{3}{5}=[0,1,1,2]$ given in Table~\ref{tab:table3}.  If the degeneracy would be $8$ it is possible to 
expand the fraction to eight components as $[0,1,1,2]=[0,1,1,2,1,-1,1,-1]$. It is still
possible if degeneracy is an odd number like $7$ then  $[0,1,1,2]=[0,1,1,3,-1,1,-1]$. 
For an arbitrary closed curve the Euler algorithm \cite{Burde1985}, (\cite{Olds1963}, pp. 24) provides values for the Berry phase.
Thus if a tangle associated to the ground state is $T =[[a_1],[a_2],[a_3],...[a_k],...[a_d]]$ with $a_1=0$ and $a_k\neq 0$. 
Then its invariant can be decomposed with the Euler algorithm as 
\begin{equation}\label{Euler}
 F(T)=a_1 + \sum\limits^d_{k>1} \frac{(-1)^k}{q_kq_{k-1}}
\end{equation}
and so 
\begin{eqnarray}\label{BerryEuler}
\frac{\gamma_{1}}{2\pi} &=& d a_1 =0 ~,\\
\frac{\gamma_{k>1}}{2\pi}&=& d \frac{(-1)^k}{q_kq_{k-1}}~,
\end{eqnarray}
where 
\begin{eqnarray}
(-1)^k &=& p_kq_{k-1} - p_{k-1}q_{k}\\
p_k &=& a_kp_{k-1} + p_{k-2}~,\\
q_k &=& a_kq_{k-1} + q_{k-2}~,
\end{eqnarray}
within initial values
\begin{eqnarray}
p_0= 1 &  &  p_{-1}= 0 ~,\\
q_0= 1 &  &  q_{-1}= 1 ~,\\
p_d = p &  & q_d = Q ~.
\end{eqnarray} 
As a  conclusion, from \eqref{BerryEuler} is deduced that there might be an state in the LLL  such that the Berry phase $\gamma_1=0$
and so the  Berry curvature vanish \cite{Sakurai}. 
This state is associated to the tangle $[0]$, with fraction F(T)=0, which produces
a perfect insulator with ultrahigh resistance. Of course this is an extreme case
where the Euler algorithm breaks down since one would be obligated to do $a_k=0$ too. Moreover
the Kubo's formula given in \eqref{Conductivity} might be invalid and apply only
for liquid like states with no vanishing fillings \cite{Thouless1985}. 
The Kauffman equation given by  \eqref{FractionEq} is consistent with a description 
based on Kubo's formula since it is forbidden  to divide by the tangle $T=[0]$.



\section{Isotopy and Potential Energy} \label{section:energy}
There is other way to obtain filling factors with torus knots.
In this section it is shown that the Laughlin explanation for the IQHE in ref. \cite{LaughlinPRB1981} 
is generalized for the case of FQHE. A beautiful relation 
emerges when isotopy between torus knots
is considered. Here the Berry connection is 
multiplied with the magnetic flux quantum $\phi_0= \frac{h}{2e}$ and added to the vector potential inside the N-particles Hamiltonian.

\subsection{Case IQHE} \label{section:IQHE}

The IQHE can be described by the  $N-$electrons Hamiltonian 
\begin{equation}\label{Hamiltonian}
 H = \sum\limits^N_{s=1} \frac{1}{2m_b}\left(p_s + \frac{e}{c} A(x_s,y_s) + \frac{e}{c} \frac{\phi_0 L}{\sqrt{N}} A\right)^2 ~.
\end{equation}
A similar approach has been performed with the help of fluxes in ref.  \cite{JainBook2007} or a vector field 
in ref. \cite{LaughlinPRB1981}. However here our vector field $A$  is the Berry connection associated with
the non-degenerated ground state. Here $A(x_s,y_s)$ is the vector potential produced by the magnetic field perpendicular to 
a 2D square sample of length $L$.  Furthermore 
the torus knot in the 2D parameter space is taken as the knot numerator of a rational tangle as $\Gamma = {\bf N}(T)$ 
and so
\begin{eqnarray}\label{NewBerryPhase}
\gamma(N(T)) &=& \oint\limits_{{\bf N}(T)} i~\langle \phi |\nabla_R| \phi\rangle\cdot dR~,
\end{eqnarray} 
Since the radious of the cyclotron orbit for the state $\phi$ depends on the magnetic field as  $r_s \sim \frac{1}{B}$, 
it is expected that under an isotopy transformation on $N(T)$ the cyclotron orbit change.
An isotopy is realized by setting a new constant value for the magnetic field $B'$. 
This relates with the old as $B'=\frac{B}{\delta}$. It is a dilatation for $\delta<1$ if the cyclotron radius is
smaller than the previous value. It is a elongation for $\delta>1$ if a cyclotron radius is
bigger than before. This isotopy 
is generally a deformation because of the change in the knot size. Therefore if the separation
between two charge carries at a given time is $r_s=R-R'$ then under a deformation it transforms
as 
\begin{equation}\label{rk}
r'_s=\delta r_s~,
\end{equation}
therefore the new knot determined by the vector $R'=\delta R$
is such that ${\bf N}(T') = \delta {\bf N}(T)$. This deformation changes the value of the filling factor and the Berry phase hence
\begin{eqnarray}\label{NewBerryPhase}
\gamma({\bf N}(T')) &=& \oint\limits_{{\bf N}(T')} i~\langle \phi |\nabla_{R'}| \phi\rangle\cdot dR'~,
\nonumber\\
&=& \oint\limits_{\delta {\bf N}(T)} i~\langle \phi |\frac{1}{\delta}\nabla_R| \phi\rangle \cdot d(\delta R)~,
\nonumber\\
&=& \oint\limits_{\delta {\bf N}(T)} i~\langle \phi |\nabla_R| \phi\rangle \cdot dR~,
\nonumber\\
&\neq & \gamma({\bf N}(T))~. 
\end{eqnarray} 
and the Berry connection is deformed as
\begin{eqnarray}\label{NewBerryConnection}
A'=i~\langle \phi |\frac{1}{\delta}\nabla_R| \phi\rangle~.
\end{eqnarray} 
As a result the N-particles Hamiltonian  is deformed as
\begin{eqnarray}\label{Hamiltonian}
 H' &=& \sum\limits^N_{s=1} \frac{1}{2m_b}\left(p_s + \frac{e}{c} A(x_s,y_s) + \frac{e}{c} \frac{\phi_0 L}{\sqrt{N}} A'\right)^2 ~,
\end{eqnarray}
this Hamiltonian induced a current  operator \cite{LaughlinPRB1981,Yong2006}. The result of considering
isotopy is 
\begin{eqnarray}\label{Current}
I=\frac{dH'}{dA'}= \sum\limits^N_{s=1}\frac{e \phi_0 L }{c\sqrt{N}  m_b}\left(p_s + \frac{e}{c} A(x_s,y_s)+ \frac{e}{c} \frac{\phi_0 L}{\sqrt{N}} A'\right)~.
\end{eqnarray} 
The average of \eqref{Current} together with an integration along the knot yields 
\begin{eqnarray}\label{IntegrationIQHE}
\frac{c^2  m_b}{2\pi h (\phi_0 L)^2} \oint\limits_{{\bf N}(T)} \langle \phi |I| \phi\rangle  \cdot dR &=&\frac{e^2}{2\pi h}\oint\limits_{{\bf N}(T)}  A' \cdot dR ~,
\nonumber\\
&=& \frac{e^2}{h}  \frac{\gamma({\bf N}(T))}{2\pi}~,
\nonumber\\
&=& \bold{n}~,
\end{eqnarray} 
which corresponds to the filling factor for the IQHE. 

\subsection{Case FQHE} \label{section:FQHE}
The Hamiltonian in the FQHE is
\begin{equation}
 {\bf H} = H +V
\end{equation}
where $V = \sum\limits^N_{i,j=1}\frac{e^2}{|\bold{r}_i-\bold{r}_j|}$ is the potential energy. 
However the Berry connection for degenerated N-particles is $A_k$, so the index $k$ labels
degenration of the ground state. Now let us define a new vector named $B_k$ such that the Berry connection obeys $A_k\cdot B_k=1$. 
If the 2D connection  is $A_k=(a_1,a_2)$ then the new vector is given by
$B_k=\frac{1}{2}(a^{-1}_1,a^{-1}_2)$. As a consequence  the Berry connection  is deformed as in  
\eqref{NewBerryConnection} by the value 
\begin{equation}\label{delta}
\delta=A_k\cdot B'_k~.
\end{equation}
Additionally notice that  $(dA_k)\cdot B_k+A_k\cdot dB_k=0$ and so
\begin{equation}\label{dBkdAk}
\frac{dB_k}{dA_k}=-\frac{1}{|A_k|^2}~.
\end{equation}
Besides the Hamiltonian transformed after an isotopy deformation is
 \begin{equation}
 {\bf H}' = H' +V'~,
\end{equation}
where the new value for the potential energy is
originated by the isotopy \eqref{rk} as $V'=\frac{V}{\delta}$.
Similar as the procedure used to obtain \eqref{Current}, the current for the FQHE  is then calculated from ${\bf I}= \frac{d{\bf H}'}{dA'_k}$
in the expression
\begin{eqnarray}\label{CurrentFQHE}
{\bf I}&=& I + \frac{dV'}{dA'_k} ~.
\end{eqnarray} 
The IQHE is recovered from \eqref{CurrentFQHE} in the absence of isotopy. 
In other words, when  $\delta=1$ and so $V'$ would equal $V$  which is independent of $A'_k$.
Furthermore the degeneration of the ground state is introduced with the standard sum $\frac{1}{d}\sum\limits^d_{k=1}$.
The average and integration of \eqref{CurrentFQHE} yields the FQHE filling factor  as
\begin{eqnarray}\label{IntegrationFQHE}
 \frac{c^2 m_b}{2\pi h (\phi_0 L)^2}\frac{1}{d}\sum\limits^d_{k=1} \oint\limits_{{\bf N}(T')}\langle \phi_k |{\bf I}| \phi_k\rangle\cdot dR = &&
\nonumber\\
\bold{n}+ \frac{c^2  m_b}{2\pi h (\phi_0 L)^2}\frac{1}{d}\sum\limits^d_{k=1}\oint\limits_{{\bf N}(T')} \langle\phi_k| \frac{dV'}{dA'_k}| \phi_k\rangle  \cdot dR~,
 \end{eqnarray} 
This gives the value for the filling factor of FQHE as 
\begin{eqnarray}\label{nuFQHEISo}
\nu=\frac{c^2 m_b}{2\pi h (\phi_0 L)^2}\frac{1}{d}\sum\limits^d_{k=1} \oint\limits_{{\bf N}(T')}\langle \phi_k |{\bf I}| \phi_k\rangle\cdot dR~,
\end{eqnarray} 
where $\bold{n}$ is calculated from \eqref{IntegrationIQHE} and it is a topological invariant of the tangle $T$.
On the other hand the filling factor $\nu$ is a topological invariant of the tangle $T'$.
It must  be extressed that \eqref{IntegrationFQHE}  can not be used as a  relation between fillings factors of IQHE  
since knot numerators of IQHE are not isotopic, see Table~\ref{tab:table0}.
For unoriented or oriented knots the expression \eqref{nuFQHEISo} equals the equations \eqref{formula2}  
and \eqref{formula3} repectively. For instance, as it was mentioned before, the value $p\to \infty$, in \eqref{formula2},
is a superconductive state which for unoriented knot generates finite filling factors for the FQHE as
\begin{equation}
\nu= \frac{1}{\alpha}~,
\end{equation}
thus from \eqref{IntegrationFQHE} and \eqref{nuFQHEISo} is obtain
\begin{equation}\label{MainResult}
\frac{c^2 m_b}{2\pi h (\phi_0 L)^2}\frac{1}{d}\sum\limits^d_{k=1}\oint\limits_{{\bf N}(T')} \langle\phi_k| \frac{dV'}{dA'_k}| \phi_k\rangle  \cdot dR = \frac{1}{\alpha}-\bold{n} ~,
\end{equation}
Consequently the potential is stable for the value $\alpha=\frac{1}{\bold{n}}$. Let us consider now the case when 
$\bold{n}$ is finite. So large values of $\nu$ are produced by innestabilities in the potential $V'$. Now the derivative
inside \eqref{MainResult} for $\delta\neq 1$ is calculated using \eqref{dBkdAk} as follows
\begin{eqnarray}\label{dVdA}
\frac{dV'}{dA'_k}&=& -\frac{V}{\delta^2} \frac{\partial\delta}{\partial A'_k}=
-\frac{V}{\delta^2} \frac{\partial(A_k\cdot B'_k)}{\partial A'_k}
= -\frac{V}{\delta^2} \left(A_k \frac{\partial B'_k}{\partial A'_k}\right)~,
\nonumber\\
&=& \frac{V}{\delta^2}  \frac{A_k}{|A'_k|^2}= V\frac{A_k}{|A_k|^2}~.
\end{eqnarray} 
Finally by replacing  \eqref{dVdA} in \eqref{MainResult} a relation between filling factors and Coulomb potential
arises as
\begin{eqnarray}\label{kVk}
\frac{c^2 m_b}{2\pi h (\phi_0 L)^2}\frac{1}{d}\sum\limits^d_{k=1} \langle\phi_k|V| \phi_k\rangle \oint\limits_{{\bf N}(T')} \frac{A_k}{|A_k|^2} \cdot dR = \frac{1}{\alpha}-\bold{n} ~.
\end{eqnarray}

\begin{acknowledgments}
We wish to acknowledge  support from the Polish National Center of Science Project no.: UMO-2011/02/A/ST3/00116.
\end{acknowledgments}

\appendix*

\section{Filling Factors of FQHE with the Tangle Method}\label{apendix}

As we show in Fig.~\ref{fig5} the tangle method can be used to classify filling factors of FQHE. Each fraction is a topological
invariant of a rational tangle. The Theorem 2 is used to organized these fractions according the isotopy of 
knot numerators. Thus using the method  exemplified in Section~\ref{section:method} one can get the tables 
Tables~\ref{tab:table1} - \ref{tab:table10}. In these tables the filling factors are confirmed by experimental results published
in ref. \cite{Eisenstein2000,Stormer2003}

\begin{table*}[!htbp]
\caption{\label{tab:table0}Filling factors for the IQHE. Notice that the knot numerators 
of tangles $[1]$ and $[\infty]$ are isotopic. Alternatings knots with more than ten crossings like $K11a364$ are denoted with letter K. 
Links start with letter L like $L2a1$ which is the Hopf-Link. The $0_2$
stays for two unknots.}
\begin{ruledtabular}
\begin{tabular}{lccccccccccc}
$\nu$&0       &1      & 2       & 3      & 4       & 5       & 6     & 7       &  p-odd &  p-even & $\infty$\\
\hline
T    & [0]    & [1]   & [2]     &  [3]   & [4]     &  [5]   & [6]    &  [7]    & [p]    & [p]     & $[\infty]$\\
N(T) & $0_2$  & $0_1$ & $L2a1$ &  $3_1$ & $L4a1$ &  $5_1$ &  L6a3  & $7_1$   &  Knot  & Link    & $0_1$\\
D(T) & $0_1$  & $0_1$ & $0_1$   &  $0_1$ & $0_1$   &  $0_1$ &  $0_1$ & $0_1$   &  $0_1$ & $0_1$   & $0_2$\\
\end{tabular}
\end{ruledtabular}
\end{table*}

\begin{table*}[!htbp]
\caption{\label{tab:table1}Filling factors for the FQHE $\frac{1}{q}$. Notice that absolutely all  knot numerators of tangles [[0],[q]]
 are isotopic to the knot $N([1])\sim0_1$ which associates with the LLL of IQHE.}
\begin{ruledtabular}
\begin{tabular}{lcccccc}
$\nu$& $\frac{1}{2}$      &$\frac{1}{3}$     & $\frac{1}{4}$      & $\frac{1}{9}$     & $\frac{1}{q}$ ($q-$odd)      & $\frac{1}{q}$ ($q-$even)       \\
\hline
T    & $[[0],[2]]$        & $[[0],[3]]$      & $[[0],[4]]$        &  $[[0],[9]]$      & $[[0],[q]]$                 &  $[[0],[q]]$   \\
N(T) & $0_1$              & $0_1$            & $0_1$              &  $0_1$            & $0_1$                       &  $0_1$          \\
D(T) & $L2a1$            & $3_1$            & $L4a1$            &  $9_1$            & Knot                        &  Link         \\
\end{tabular}
\end{ruledtabular}
\end{table*}

\begin{table*}[!htbp]
\caption{\label{tab:table2}Filling factors for $\frac{2}{q}$. All  knot numerators of tangles [[0],[k],[2]] are isotopic to $N([2])\sim L2a1$.}
\begin{ruledtabular}
\begin{tabular}{lcccc}
$\nu$& $\frac{2}{3}$      &$\frac{2}{5}$     & $\frac{2}{7}$      & $\frac{2}{2k+1}$ ($k= 0,1,2,...$)        \\
\hline
T    & $[[0],[1],[2]]$    & $[[0],[2],[2]]$  & $[[0],[3],[2]]$    & [[0],[k],[2]]\\
N(T) & $L2a1$            & $L2a1$          & $L2a1$            & $L2a1$   \\
D(T) & $3_1$              & $4_1$            & $5_2$              & Knot    \\
\end{tabular}
\end{ruledtabular}
\end{table*}

\begin{table*}[!htbp]
\caption{\label{tab:table3}Filling factors for  $\frac{3}{q}$. Knot denominators have $k+2$ crossings. They are isotopy to $N([3])\sim3_1$}
\begin{ruledtabular}
\begin{tabular}{lcccccc}
$\nu$& $\frac{3}{2}$      &$\frac{3}{4}$     & $\frac{3}{5}$           & $\frac{3}{7}$         & $\frac{3}{8}$               & $\frac{3}{11}$       \\
\hline
T    & $[[1],[2]]$        & $[[0],[1],[3]]$  & $[[0],[1],[1],[2]]$     &  $[[0],[2],[3]]$      & $[[0],[2],[1],[2]]$         &  $[[0],[3],[1],[2]]$   \\
N(T) & $3_1$              & $3_1$            & $3_1$                   &  $3_1$                & $3_1$                       &  $3_1$          \\
D(T) & $L2a1$            & $L4a1$            & $4_1$                 &  $5_2$                & $L5a1$                     &  $6_1$         \\
\end{tabular}
\end{ruledtabular}
\end{table*}

\begin{table*}[!htbp]
\caption{\label{tab:table4}Filling factors for $\frac{4}{q}$.  Isotopy to $N([4])\sim4_1$ }
\begin{ruledtabular}
\begin{tabular}{lcccccc}
$\nu$& $\frac{4}{3}$      &$\frac{4}{5}$     & $\frac{4}{7}$           & $\frac{4}{9}$         & $\frac{4}{11}$              & $\frac{4}{15}$       \\
\hline
T    & $[[1],[3]]$        & $[[0],[1],[4]]$  & $[[0],[1],[1],[3]]$     &  $[[0],[2],[4]]$      & $[[0],[2],[1],[3]]$         &  $[[0],[3],[1],[3]]$   \\
N(T) & $L4a1$            & $L4a1$          & $L4a1$                 &  $L4a1$              & $L4a1$                     &  $L4a1$          \\
D(T) & $3_1$              & $5_1$            & $5_2$                   &  $6_1$                & $6_2$                       &  $7_4$         \\
\end{tabular}
\end{ruledtabular}
\end{table*}

\begin{table*}[!htbp]
\caption{\label{tab:table5}Filling factors for $\frac{5}{q}$. Since $4_1$  and $5_1$ are not isotopic 
one is forced to use the two options of the Theorem 2. For instance the fraction $\frac{5}{9}$
can be obtained with the first option while the fraction $\frac{5}{11}$ needs the second option.
This is a sign of unconventionallity. 
}
\begin{ruledtabular}
\begin{tabular}{lccccccc}
$\nu$& $\frac{5}{2}$      &$\frac{5}{3}$       & $\frac{5}{7}$           & $\frac{5}{9}$         & $\frac{5}{13}$              & $\frac{5}{13}$             & $\frac{5}{19}$     \\
\hline
T    & $[[2],[2]]$        & $[[1],[1],[2]]$    & $[[0],[1],[2],[2]]$     &  $[[0],[1],[1],[4]]$  & $[[0],[2],[5]]$             &  $[[0],[2],[1],[1],[2]]$   & $[[0],[3],[1],[4]]$  \\
N(T) & $4_1$              & $4_1$              & $4_1$                   &  $5_1$                & $5_1$                       &  $4_1$                     & $5_1$  \\
D(T) & $L2a1$            & $3_1$            & $5_2$                   &  $6_1$                & $6_2$                     &  $7_4$                     & $8_4$  \\
\end{tabular}
\end{ruledtabular}
\end{table*}

\begin{table*}[!htbp]
\caption{\label{tab:table6}Filling factors for $\frac{6}{q}$.}
\begin{ruledtabular}
\begin{tabular}{lcccc}
$\nu$& $\frac{6}{11}$             &$\frac{6}{13}$      & $\frac{6}{23}$           & $\frac{6}{25}$                \\
\hline
T    & $[[0],[1],[1],[5]]$        & $[[0],[2],[6]]$    & $[[0],[3],[1],[5]]$      &  $[[0],[4],[6]]$        \\
N(T) & $L6a3$                      & $L6a3$              & $L6a3$                    &  $L6a3$                         \\
D(T) & $7_2$                    & $8_1$            & $9_5$                    &  $10_3$                     \\
\end{tabular}
\end{ruledtabular}
\end{table*}

\begin{table*}[!htbp]
\caption{\label{tab:table7}Filling factors for $\frac{7}{q}$. Since $5_2$  and $7_1$ are not isotopic. Therefore
unconventionallity is present here.}
\begin{ruledtabular}
\begin{tabular}{lccccccc}
$\nu$& $\frac{7}{2}$      &$\frac{7}{3}$       & $\frac{7}{5}$           & $\frac{7}{11}$         & $\frac{7}{13}$              & $\frac{7}{15}$             & $\frac{7}{27}$     \\
\hline
T    & $[[3],[2]]$        & $[[2],[3]]$    & $[[1],[2],[2]]$     &  $[[0],[1],[1],[1],[3]]$  & $[[0],[1],[1],[6]]$             &  $[[0],[2],[7]]$   & $[[0],[3],[1],[6]]$  \\
N(T) & $5_2$              & $5_2$              & $5_2$                   &  $5_2$                & $7_1$                       &  $7_1$                     & $7_1$  \\
D(T) & $L2a1$            & $3_1$            & $4_1$                   &  $6_2$                & $8_1$                     &  $9_2$                     & $10_4$  \\
\end{tabular}
\end{ruledtabular}
\end{table*}

\begin{table*}[!htbp]
\caption{\label{tab:table8}Filling factors for $\frac{8}{q}$. Knots start with letter K like $K11a364$ which is a knot of
eleven crossings. The Hoste-Thistlethwaite table for eleven crossings can be found in \cite{Atlas}. }
\begin{ruledtabular}
\begin{tabular}{lccc}
$\nu$& $\frac{8}{15}$             &$\frac{8}{17}$      & $\frac{8}{31}$                           \\
\hline
T    & $[[0],[1],[1],[7]]$        & $[[0],[2],[8]]$    & $[[0],[3],[1],[7]]$             \\
N(T) & $L8a14$                 & $L8a14$         & $L8a14$                                          \\
D(T) & $9_2$                      & $L10a114$              & $K11a364$                                        \\
\end{tabular}
\end{ruledtabular}
\end{table*}

\begin{table*}[!htbp]
\caption{\label{tab:table9}Filling factors for $\frac{9}{q}$. Since $6_1$  and $9_1$ are not isotopic. Therefore
unconventionallity is present here. $K12a$  denotes one of the
alternating knots with twelve crossings \cite{TableKnots}.}
\begin{ruledtabular}
\begin{tabular}{lccc}
$\nu$& $\frac{9}{13}$             &$\frac{9}{19}$      & $\frac{9}{35}$                           \\
\hline
T    & $[[0],[1],[2],[4]]$        & $[[0],[2],[9]]$    & $[[0],[3],[1],[8]]$             \\
N(T) & $6_1$                      & $9_1$              & $9_1$                                           \\
D(T) & $7_3$                    & $K11a247$            & $K12a$                                        \\
\end{tabular}
\end{ruledtabular}
\end{table*}

\begin{table*}[!htbp]
\caption{\label{tab:table10}Filling factors for $\frac{10}{q}$.  $K13a$ denotes 
alternating knots with thirteen crossings \cite{TableKnots}. }
\begin{ruledtabular}
\begin{tabular}{lccc}
$\nu$& $\frac{10}{21}$             &$\frac{10}{19}$      & $\frac{10}{39}$                           \\
\hline
T    & $[[0],[2],[10]]$        & $[[0],[1],[1],[9]]$    & $[[0],[3],[1],[9]]$             \\
N(T) & $10_1$                      & $10_1$             & $10_1$                                           \\
D(T) & $K12a$                      & $K11a247$             & $K13a$                                        \\
\end{tabular}
\end{ruledtabular}
\end{table*}








\newpage 
\bibliography{apssamp}

\end{document}